\documentclass[sigconf,nonacm]{acmart}

\usepackage{comment}
\usepackage{xspace}
\usepackage{mdframed}
\usepackage{amsmath}

\usepackage{amssymb}
\usepackage{tabularx}
\usepackage{graphicx}  
\usepackage{subcaption} 
\usepackage{pgfplots}
\usepackage{pgfplotstable}
\AtBeginDocument{%
  }

\setcopyright{acmlicensed}
\copyrightyear{2018}
\acmYear{2018}
\acmDOI{XXXXXXX.XXXXXXX}
\acmConference[Published]{Make sure to enter the correct
  conference title from your rights confirmation email}{Sept 03, 2025}{Woodstock, NY}
\acmISBN{978-1-4503-XXXX-X/2018/06}

\begin{document}
\title{Efficient Item ID Generation for Large-Scale LLM-based Recommendation}

\author{Anushya Subbiah}
\orcid{0009-0009-6005-6050}
\affiliation{%
  \institution{Google Research}
  \city{Mountain View}
  \country{USA}
}
\email{anushyas@google.com}

\author{Vikram Aggarwal}
\orcid{0009-0004-1750-6806}
\affiliation{%
  \institution{Google Research}
  \city{Mountain View}
  \country{USA}
}
\email{viki@google.com}

\author{James Pine}
\orcid{0009-0006-9517-2906}
\affiliation{%
  \institution{Google Research}
  \city{Mountain View}
  \country{USA}
}
\email{rubicon@google.com}

\author{Steffen Rendle}
\orcid{0009-0004-6389-2509}
\affiliation{%
  \institution{Google Research}
  \city{Mountain View}
  \country{USA}
}
\email{srendle@google.com}

\author{Krishna Sayana}
\orcid{0009-0001-6138-0874}
\affiliation{%
  \institution{Google Research}
  \city{Mountain View}
  \country{USA}
}
\email{ksayana@google.com}

\author{Kun Su}
\orcid{0009-0004-8112-9419}
\affiliation{%
  \institution{Google Research}
  \city{Mountain View}
  \country{USA}
}
\email{sukun@google.com}

\renewcommand{\shortauthors}{Anushya Subbiah et al.}

\newcommand{\generativemodel}{generative model\xspace}
\newcommand{\generativemodels}{generative models\xspace}
\newcommand{\languagemodel}{language model\xspace}
\newcommand{\languagemodels}{language models\xspace}
\newcommand{\LLM}{LLM\xspace}
\newcommand{\LLMs}{LLMs\xspace}
\newcommand{\rs}{RS\xspace}
\newcommand{\recommendermodel}{recommender model\xspace}
\newcommand{\recommendermodels}{recommender models\xspace}
\newcommand{\recommender}{recommender\xspace}
\newcommand{\recommenders}{recommenders\xspace}
\newcommand{\itemid}{item ID\xspace}
\newcommand{\itemids}{item IDs\xspace}
\newcommand{\singlestep}{single-step\xspace}
\newcommand{\Singlestep}{Single-step\xspace}
\newcommand{\multistep}{multi-step\xspace}
\newcommand{\singletoken}{single-token\xspace}
\newcommand{\Singletoken}{Single-token\xspace}
\newcommand{\multitoken}{multi-token\xspace}
\newcommand{\prefill}{prefill\xspace}
\newcommand{\decode}{decode\xspace}
\newcommand{\ascomment}[1]{}
\newcommand{\srcomment}[1]{}
\newcommand{\argmax}{\mathop{\textrm{argmax}}}
\newcommand{\setI}{\mathcal{I}}
\newcommand{\setV}{\mathcal{V}}
\newcommand{\setEV}{E^{V}}
\newcommand{\setEC}{E^{C}}
\newcommand{\setEI}{E^{I}}
\newcommand{\setU}{\mathcal{U}}
\newcommand{\setC}{\mathcal{C}}
\renewcommand{\O}{\mathcal{O}}


\begin{abstract}
    Integrating product catalogs and user behavior into LLMs can enhance recommendations with broad world knowledge, but the scale of real-world item catalogs, often containing millions of discrete item identifiers (Item IDs), poses a significant challenge. This contrasts with the smaller, tokenized text vocabularies typically used in LLMs. The predominant view within the LLM-based recommendation literature is that it is infeasible to treat item ids as a first class citizen in the LLM and instead some sort of tokenization of an item into multiple tokens is required. However, this creates a key practical bottleneck in serving these models for real-time low-latency applications.

Our paper challenges this predominant practice and integrates item ids as first class citizens into the LLM. We provide simple, yet highly effective, novel training and inference modifications that enable \textit{\singletoken} representations of items and \textit{\singlestep} decoding. Our method shows improvements in recommendation quality (Recall and NDCG) over existing techniques on the Amazon shopping datasets while significantly improving inference efficiency by 5x-14x.
Our work offers an efficiency perspective distinct from that of other popular approaches within LLM-based recommendation, potentially inspiring further research and opening up a new direction for integrating IDs into LLMs. Our code is available here \url{https://drive.google.com/file/d/1cUMj37rV0Z1bCWMdhQ6i4q4eTRQLURtC}
\end{abstract}



\maketitle

\begin{table}[h]
    \centering
    \begin{tabular}{ll}
    \toprule
        Symbol& Description  \\
    \midrule
        $\setV$ & vocabulary of text tokens (from the \LLM) \\
        $\setI$ & catalog of all items \\
        $\setC$ & set of clusters \\
        $x \in \setV$ & text token \\
        $i \in \setI$ & item \\
        $w \in (\setV \cup \setI)$ & token \\
        $d$ & dimension of token embedding \\
        $\setEV \in \mathbb{R}^{\lvert\setV\rvert \times d}$ & embeddings of all text tokens (from the \LLM) \\
        $\setEI \in \mathbb{R}^{\lvert\setI\rvert \times d}$ & embeddings of all \itemids \\
        $\setEC \in \mathbb{R}^{\lvert\setC\rvert \times d}$  & cluster centroid embeddings of all clusters \\
    \bottomrule
    \end{tabular}
    \caption{Symbols}
    \label{tab:symbols}
\end{table}
\ascomment{item token, text token, token. naming}

\begin{figure*}[t]
    \centering
    \begin{subfigure}[b]{0.56\linewidth}
        \centering
        \includegraphics[width=\linewidth]{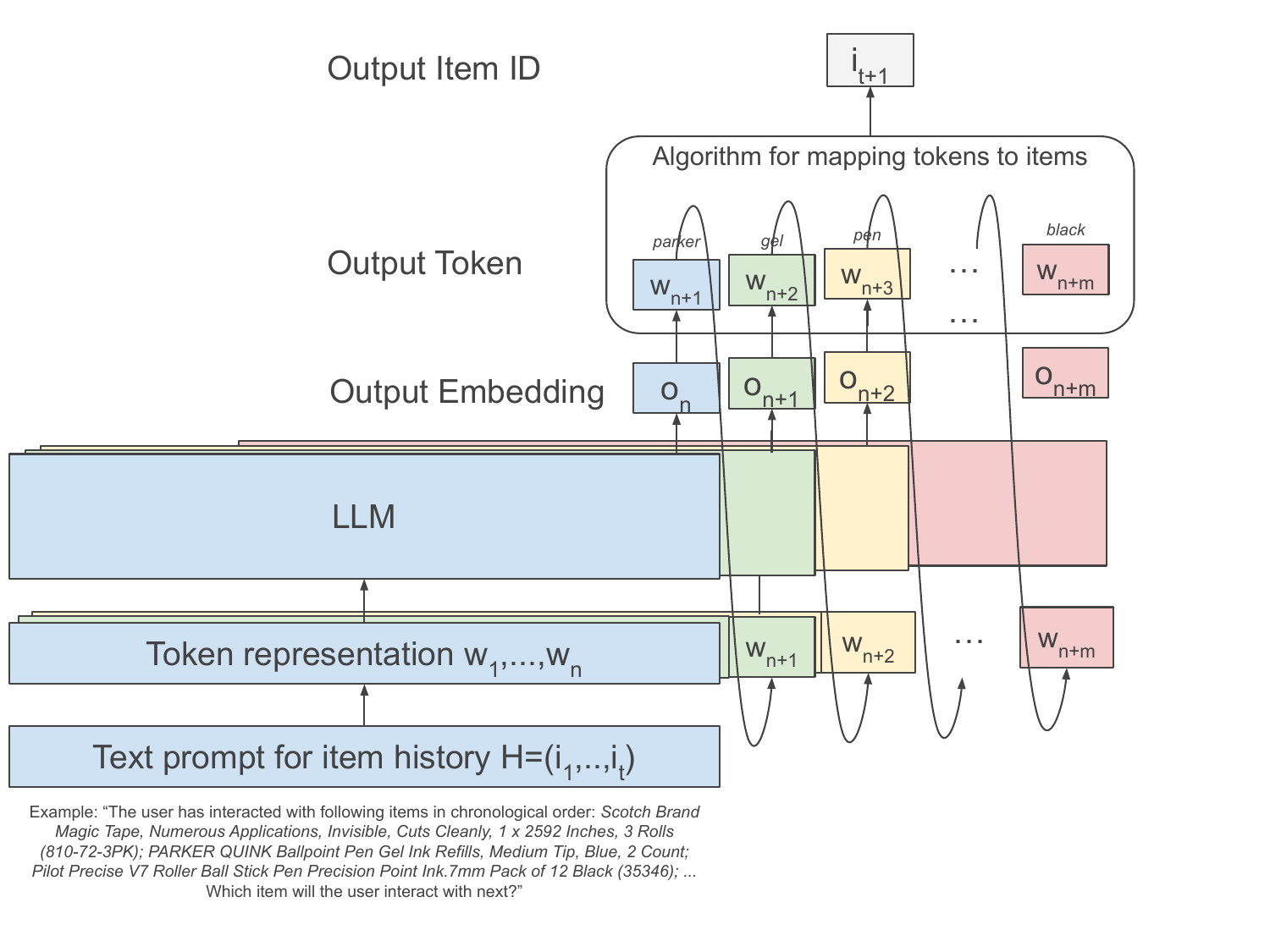}
        \caption{Multi-token representation (Baselines): Total latency is $l_{\text{prefill}}(m \cdot |H|+\text{const})  +  m \cdot l_{\text{decode}}$, where $m$ is the number of tokens to represent a single item.}
        \label{fig:multi_step}
    \end{subfigure}
    \hspace{10pt}
    \begin{subfigure}[b]{0.40\linewidth}
        \centering
        \includegraphics[width=\linewidth]{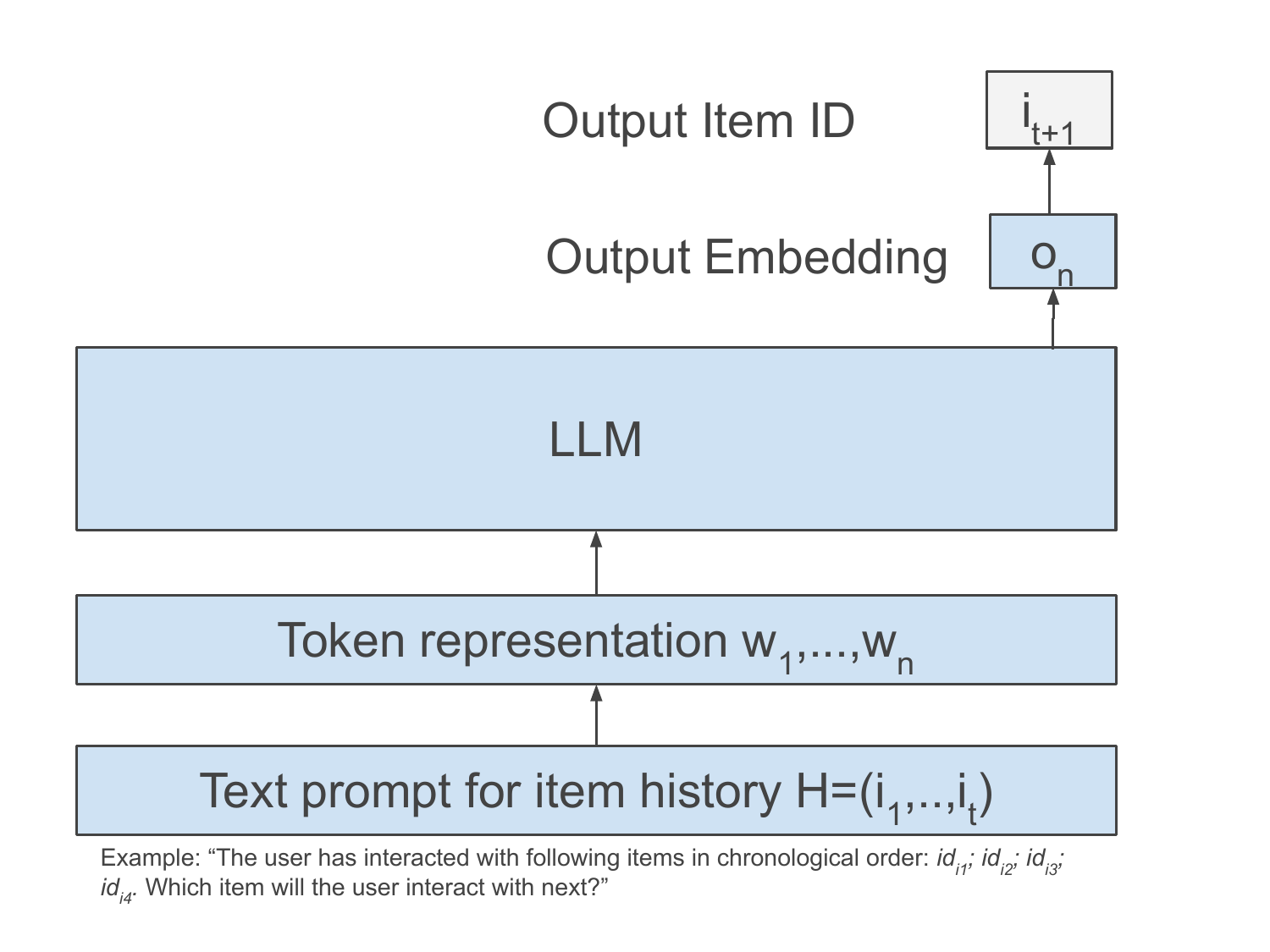}
        \vspace{1pt}
        \caption{Single-token represenation (Ours): Total latency is $l_{\text{prefill}}(|H|+\text{const})  + l_{\text{decode}}$.
        }
        \label{fig:single_step}
    \end{subfigure}

   \caption{This figure highlights the key distinction between baselines and our technique's \textit{\singletoken} representation of items in LLM-based inference. Multi-step methods require multiple iterations of the LLM, each with a latency of $l_{\text{decode}}$, while our single-step approach completes the process with a single $l_{\text{decode}}$. Both methods incur an initial latency of $l_{\text{prefill}}(m\cdot |H| + \text{const})$ for parallel input processing and this latency is also lower as $m=1$ in our case.}
    \label{fig:combined_figure}
\end{figure*}

\section{Introduction}
\label{sec:introduction}

Recommender systems (\rs) are crucial for online platforms, helping users navigate vast item catalogs~\cite{covington2016deep}. Traditional \rs often rely on collaborative filtering techniques like matrix factorization~\cite{koren2009matrix} and two-tower models~\cite{yi2019sampling} to capture user-item interactions. These methods learn latent representations for users and items, with the item embedding matrix scaling linearly with the catalog size ($|\mathcal{I}|$). While effective, these approaches may not fully utilize the rich semantic information in item descriptions and user preferences. Integrating \LLMs into \rs promises to combine collaborative filtering with semantic understanding, potentially leading to more accurate and context-aware recommendations.

\textbf{The challenge of inference latency in \LLM-based recommendation:} Most existing approaches~\cite{zhu_beyond_2025, CALRec, recformer, P5, harteleveraging2023} rely on multiple tokens (\multitoken) to represent an item by forcing items through the standard text tokenization process of LLMs. However, a critical challenge hinders the practical deployment of these \LLM-based \recommenders: the inference latency associated with generating recommendations. They generate textual descriptions or tokenized representations of items, requiring a longer \prefill input and multiple sequential calls to the \LLM to \decode each recommendation. This design fundamentally limits real-time performance, dramatically increasing inference latency and posing a severe bottleneck in real-time scenarios such as e-commerce, dynamic ad placement, and potentially even slate decoding(set of items to present to a user, considering the collective impact of those items).

\textbf{Why is \prefill and \decode a bottleneck?} LLMs generate text auto-regressively, meaning each new token depends on the previously generated ones. This process has two phases: a parallel \textit{\prefill} phase that processes the input prompt, and a sequential \textit{\decode} phase where each token is generated one by one, each requiring a full forward pass through the model \cite{popeefficiently2022, agrawal2024taming, shazeerfastdecoding2019}. Current LLM-based recommenders often use multi-step decoding and longer prefill input, representing each item as a sequence of tokens (see Figure~\ref{fig:multi_step}). In real-world scenarios inputs are dominated by potentially hundreds of user history \itemids and the number of items is often O(millions). Further, prefill and decode latency increase with \LLM size and as we will see later in our experiments, recommendation quality increases with larger \LLMs, emphasizing the importance of our efficient method.

\textbf{Vocabulary disparity and computational cost} LLMs and traditional recommender models differ substantially in their output spaces. LLMs utilize tokenizers like WordPiece~\cite{devlin2019bert}, resulting in a vocabulary ($|\setV|$) of only thousands of text tokens. Recommender systems, however, must handle item catalogs ($|\setI|$) with potentially millions of discrete \itemids. This disparity creates a computational bottleneck in the LLM's output layer. The standard softmax requires a matrix multiplication with time and space complexity $O((|\setV| + |\setI|) \cdot d)$ per example, where $d$ is the embedding dimension. While a significant portion of LLM parameters and research effort are dedicated to the attention mechanism (with its quadratic time complexity in input sequence length, as addressed by methods like Reformer~\cite{kitaev2020reformer}), our work tackles the distinct, and for large-scale recommendation settings often dominant bottleneck of the softmax output layer and item input length.

\textbf{Existing approaches and their limitations} Many techniques have been proposed to incorporate \LLMs into recommendation systems. One common method is to represent items in the model input and output using their textual metadata, such as titles, descriptions, or categories~\cite{CALRec, recformer, P5, harteleveraging2023}. Another strategy involves tokenizing \itemids into multiple sub-item tokens, either directly~\cite{huaindexing} or using hash tables derived from separately trained models~\cite{zhu_beyond_2025, tiger2023}. These methods are aimed to reduce the softmax vocabulary size but still rely on \textit{\multistep} decoding, where each individual step within that multi-step process requires a full forward pass of the \LLM. Also, the user history is the majority of input prefill and hence prefill length scales proportional to the number of sub-item tokens. This makes the generation of items computationally expensive in latency-sensitive scenarios. While this leverages the natural language processing and generation capabilities of \LLMs, it introduces challenges during inference, as accurately mapping the generated tokens back to specific \itemids can add additional complication. To mitigate this, most existing techniques also use specialized approximate beam search (e.g., in CALRec~\cite{CALRec}). These limitations highlight the need for a more efficient approach to practically applying LLMs in recommendation systems where low-latency is crucial.

This paper tackles the scalability challenges of using LLMs for recommendation with large item catalogs and long user interaction history. \textbf{Our contributions}:

\begin{itemize}
\item Identifying the significant prefill-decode bottleneck caused by predominant techniques in LLM-based recommendation literature, and \textbf{fundamentally challenging} the prevailing assumption that items must be mapped into the text token space. We identify and demonstrate that representing items via single, direct embeddings offers a superior strategy for LLM-based recommenders.
\item Training and inference optimizations, including a Two-Level softmax based on precomputed clusters, that enable efficient training and, crucially, \textbf{efficient single-step decoding} for low-latency recommendations using two methods: Inference using Hierarchical Structure and Inference using Approximate Nearest Neighbor (ANN) search.
\item A \textbf{novel and crucial training paradigm} that leverages rich textual item metadata during training while enabling \textbf{highly efficient, ID-only inference}. Our simpler, end-to-end single-phase training achieves higher quality than multi-stage methods predominant in our baselines and in LLM-based recommendation domain.
\end{itemize}

Figure~\ref{fig:combined_figure} provides a visual overview of our approach.

\section{Related Work}
\label{sec:related_work}

Our work builds upon two main research areas: sequential item prediction and the application of \LLMs to \rs.

\subsection{Sequential Item Prediction}

Sequential item prediction aims to forecast a user's next item of interest based on their past interactions. Early approaches, such as GRU4Rec~\cite{gru4rec}, employed Recurrent Neural Networks (RNNs), specifically Gated Recurrent Units (GRUs), to model sequential dependencies in user behavior. Subsequent research explored the use of self-attention mechanisms. For instance, AttRec~\cite{zhang2019next} models user intent within a session using self-attention and incorporates personalization through metric learning for user-item affinity. SASRec~\cite{kang2018self}, utilizes self-attention to capture long-range dependencies in user sequences.

More recently, inspired by the success of masked language modeling in natural language processing, BERT4Rec~\cite{sun2019bert4rec} and S3-Rec~\cite{s3rec} adapted Transformer models with masking strategies for sequential recommendation. While effective, these methods primarily rely on behavioral data and may not fully leverage the semantic information present in item descriptions and metadata.

More recent works have started exploring generative retrieval methods~\cite{2024genrecsys}. These methods leverage the capabilities of generative models to directly predict the next item, which is the area our work falls under.

\subsection{\LLMs for Recommendation}
The field of recommender systems has witnessed a surge of interest in leveraging the language understanding and generation capabilities of \LLMs. Several approaches have been proposed to incorporate user behavior and textual item information into \LLMs for recommendation.

One category uses variable-length textual representations for items, leveraging the LLM's natural language processing abilities. Methods like CALRec~\cite{CALRec}, Recformer~\cite{recformer} and GPT4Rec~\cite{gpt4rec} frame recommendation as a natural language task, using item titles or other metadata. This is beneficial for domains with rich textual item information or limited behavioral data. Another category ~\cite{zhu_beyond_2025, huaindexing, differentiableindex2022, tiger2023, yang2022openp5} employs fixed-length tokenized representations learned from behavioral or semantic data. Crucially, these approaches typically rely on \multitoken representation of items in input user history and decoding of multiple tokens to generate an item. However, accurately mapping generated tokens back to specific item IDs can be challenging, often requiring adapted beam search (e.g., CALRec \cite{CALRec}). The exact number of steps and the cost of LLM varies, but each step is sequential and incurs an LLM forward pass. In contrast, our work focuses on novel training and inference optimizations to enable efficient prefill-decode inference latency and adapt the LLM's output mechanism to efficiently handle large item catalogs, as detailed in Section \ref{sec:method}.

Several existing works tackle training efficiency techniques such as frozen item parameters, LoRA\cite{lora}. These efficient training techniques are orthogonal to our approach as our goal is efficient inference.

\section{Problem Setting}
\label{sec:problem_setting}

Refer to Table~\ref{tab:symbols} for definitions of the symbols used throughout the paper. Let $\setI$ denote the set of all items that can be recommended, referred to as the \emph{item catalog}. Each item $i \in \setI$ is associated with an \itemid and potentially with additional textual metadata (e.g., title, brand).

We consider implicit feedback in the form of past positive interactions between users and items, such as clicks, purchases, or views. For a given user, we represent their interaction history as a chronologically ordered sequence $H = (i_1, i_2, ..., i_t)$, where each $i_j \in \setI$ is the item interacted with at time step $j$.

Our goal is to build a generative recommender system that can predict the next item a user will interact with, given their interaction history. Formally, we aim to model the conditional probability distribution $P(i_{t+1} | H)$, where $i_{t+1} \in \setI$ is the next item to be recommended at time step $t+1$.

\subsection{Item Recommendation with an LLM}
\label{sec:task_definition}

We frame the sequential item prediction task as an autoregressive language modeling problem.
We construct the input sequence for a user history $H$ by interleaving text and item tokens. An example of the LLM prompt is shown below:
\newmdenv[
  backgroundcolor=lightgray!30,
  linecolor=black,
  linewidth=1pt,
  innertopmargin=5pt,
  innerbottommargin=5pt,
  innerleftmargin=10pt,
  innerrightmargin=10pt,
  roundcorner=5pt,
  skipabove=5pt,
  skipbelow=5pt,
]{highlightbox}

\begin{highlightbox}
The user has interacted with following items in chronological order 
id: $\mathrm{id}_1$ title: $\mathrm{title}_1$, brand: $\mathrm{brand}_1$, price: $\mathrm{price}_1$,
category: $\mathrm{category}_1$, id: $\mathrm{id}_2$ .., ...., id: $\mathrm{id}_t$ .. category: $\mathrm{category}_t$.
Which item will the user interact with next? id: $\mathrm{id}_{t+1}$
\end{highlightbox}

The model is trained to efficiently generate the item $\mathrm{id}_{t+1}$ following the prompt "Which item will the user interact with next? id:". We leave other specifics of the training task to Section~\ref{sec:training_paradigm} and Section~\ref{sec:experiments}. Our proposed method is general and can be adapted to other recommendation tasks that can be formulated as text generation problems. For example, the input could be modified to include user review information or the output can be modified to predict a multi-modal representation of item.

\subsection{Input Representation of the LLM}
\label{sec:input_representation}

We shortly describe how the prompt of a history $H$ with $t$ items is mapped to a tokenized input representation with $n$ tokens ($w_1, \ldots, w_n$) where each token $w_j$  can be a text token from $\setV$ or an item token from $\setI$ (see figure~\ref{fig:combined_figure}).

For the textual components (e.g., item metadata, user instructions), we follow the standard \LLM approach: We use the pre-trained \LLM's tokenizer to convert the text into a sequence of tokens from the vocabulary $\setV$, which are then mapped to their corresponding embeddings from the \LLM's embedding matrix $\setEV$.

For the \itemid components, we store a separate item embedding table of size $\lvert\setI\rvert \times k$. To align the item embeddings with the \LLM's embedding space, we use a small feedforward neural network (projection head) to project the $k$-dimensional item embeddings into the $d$-dimensional \LLM embedding space. We refer to the embeddings after the projection as $\setEI \in \mathbb{R}^{\lvert\setI\rvert \times d}$. We interleave the item and text embeddings to create a sequence of input embeddings to the LLM.
The item embeddings are trained along with the pre-trained \LLM on the final task.

Note that our use of the term "\singletoken" is an analogy: 1 item $\rightarrow$ 1 direct embedding vector, contrasting with "\multitoken" 1 item $\rightarrow$ multiple text token embeddings.

\begin{figure*}[ht]
    \centering
    \begin{subfigure}[b]{0.42\linewidth}
        \centering
        \includegraphics[width=\linewidth]{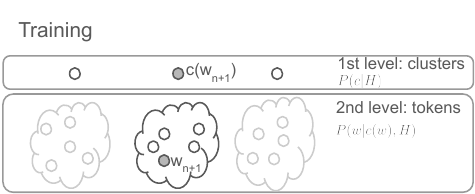}
        \caption{Training for a token $w_{n+1}$}
        \label{fig:two_level_training}
    \end{subfigure}
    \hfill
    \begin{subfigure}[b]{0.26\linewidth}
        \centering
        \includegraphics[width=\linewidth]{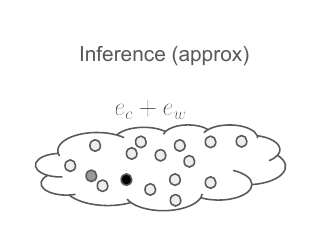}
        \caption{Inference using ANN}
        \label{fig:two_level_ann}
    \end{subfigure}
        \hfill
 \begin{subfigure}[b]{0.26\linewidth}
        \centering
        \includegraphics[width=\linewidth]{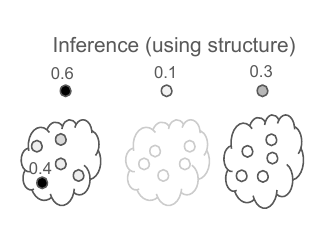}
        \caption{Inference using Structure}
        \label{fig:two_level_inference}
    \end{subfigure}
   \caption{This figure illustrates the efficiency gains of our Two-Level softmax during training and inference.
   For training, only the cluster $c(w_{n+1})$ of the target item $w_{n+1}$ is visited, all other clusters are ignored. At inference time, we leverage either ANN search or the hierarchical structure for efficient retrieval. Here, we can eliminate the second cluster with $P(c_2|H) = 0.1 < 0.6\cdot 0.4 = 0.24$.}
    \label{fig:combined_figure2}
\end{figure*}

\subsection{Decoding Items from an LLM}

The output of the last layer of the LLM at the very last position $n$ is a $d$ dimensional embedding that we denote with $o_n$.
Based on this output embedding, we want to generate an item id.
Section~\ref{sec:method} describes our single-step decoding approach.

In contrast, multi-step decoding (see figure~\ref{fig:multi_step}) would predict a token $w_{n+1}$ from $o_n$ that is appended to the existing input and the LLM would be invoked on the input $(w_{1},\ldots,w_{n},w_{n+1})$ to create an embedding $o_{n+1}$.
Multi-step decoding would repeat this process until the tokens $w_{n+1}, w_{n+2}, \ldots$ can be mapped to one of the item IDs in $\setI$.
Single-step decoding does not need this repeated process and generates the item ID directly from $o_{n}$.

\textbf{Crucially:} We map item IDs directly to item embeddings (from a dedicated embedding matrix), completely bypassing the LLM's text tokenizer. Instead of the inbuilt softmax, we use an efficient softmax to predict items directly from their own dedicated embedding vectors (see section~\ref{sec:method}). The core LLM machinery (transformer blocks) remains unchanged, operating on the embeddings provided.

\subsection{Training paradigm for ID-only inference}
\label{sec:training_paradigm}

We introduce a novel training paradigm designed to reconcile the LLM's ability to leverage rich textual item metadata (title, brand, price, category) with the necessity of efficient, ID-only inference in practical recommendation systems where inputs are dominated by potentially hundreds of user history \itemids.

During training, we randomly sample a subset of the item metadata (\itemid, title, brand, price, category) for each item in the input sequence and similarly for the label. A crucial 25\% of training examples deliberately use only the \itemid for both input and output. This specifically trains the model to depend solely on \itemids during inference for evaluation. During inference, each item in the history is represented in the input using a single item embedding (ours) instead of sequences of text token embeddings (baselines), hence our prefill is efficient.

Other LLM recommenders use domain specific pre-training stages or pre-trained item embeddings. With the training paradigm that we outlined above these multi-stage training methods could be removed and provided no improvement over just training the item embeddings from scratch. This is a further simplification over current state-of-the-art LLM recommender methods.

\section{Efficient Single-Step Item ID Generation}
\label{sec:method}

This section introduces approaches for single-step ID generation.
First, we describe full softmax which we prescribe if the item catalog is not very large. However, in real-world systems the item catalog is often O(millions) and in this case we propose a Two-Level softmax that can deal with larger catalogs. Due to the hierarchical structure of Two-Level softmax, the runtime is sublinear in the number of items and the model can be trained efficiently. We recommend this approach as opposed to using \multitoken representation of items.
Our novelty for LLM-based recommendation is the finding that \singletoken item representation is more efficient for inference and yields better quality than the dominant \multitoken approach.

\subsection{Full Softmax}
\label{subsec:fullsoftmax}

In a standard \LLM, $o_n$ is used to compute a probability distribution over text tokens $x$ from the vocabulary $\setV$ via a softmax function.
This approach can be carried over to predicting \itemids as well.
The probability of the next token (\itemid or text token) reads:
\begin{align}
    P(w|H) = \frac{\exp(\langle o_n, e_w \rangle)}{\sum_{w' \in (\setI \cup \setV)} \exp(\langle o_n, e_{w'} \rangle)}, \quad \text{where} \ w \in (\setV \cup \setI) \label{eq:fullsoftmax}
\end{align}
However, directly computing a softmax over the combined vocabulary $\setV \cup \setI$ can become computationally expensive due to the large size of $\setI$.
This is because the partition function (denominator in eq.~\ref{eq:fullsoftmax}) requires a matrix multiplication between $o_t$ and the embedding matrix of the combined vocabulary, which has dimensions $(|\setV| + |\setI|) \times d$.
Therefore, training can become impractical for large item catalogs.
It should be mentioned that inference for large item catalogs is less of an issue because ANN algorithms can find efficiently the top scoring items  even in very large item catalogs.

To summarize, as long as the item catalog is small enough for practical training, full softmax can be directly applied for single-step decoding.
In our evaluation in Section~\ref{sec:experiments}, we show that full softmax can achieve high quality item recommendations.

\begin{table*}[t]
    \centering
    \begin{tabular}{|l|c|c|}
        \hline
        \textbf{Property} & \textbf{Our Method} & \textbf{Existing Approaches (e.g., multi-token)} \\
        \hline
        Decoding Latency (per item) & $l_{\text{prefill}}(|H|+\text{const})  + l_{\text{decode}}$ & $l_{\text{prefill}}(m \cdot |H|+\text{const})  +  m \cdot l_{\text{decode}} $ \\
        \hline
        Item Representation Dimension & Flexible (e.g., $k=512$) & Constrained by \LLM's $d$ \\
        \hline
        Storage Overhead & Moderate (e.g., $0.5$ billion parameters for 1M items) & Lower (no separate item embeddings) \\
        \hline
    \end{tabular}
    \caption{Comparison of key properties between our method and existing approaches. Here $m$ represents the number of tokens that are used for representing an item (for single-token methods (ours) $m=1$), and \textmd{$\text{const}$} represents other constant prefill costs, e.g. for the prompt.}
    \label{tab:model_properties}
\end{table*}

\subsection{Two-Level Softmax}
\label{subsec:two_pass_softmax}

To overcome the computational bottleneck of full softmax for large catalogs, we take inspiration from existing research on reducing softmax complexity, such as hierarchical softmax~\cite{morin2005hierarchical}. As shown in Figure~\ref{fig:combined_figure2}, instead of directly computing a softmax over $\setV \cup \setI$, we introduce a novel Two-Level softmax mechanism specifically designed for our problem.

Our Two-Level softmax efficiently approximates the probability distribution over $\setV \cup \setI$ by introducing a hierarchical structure. We modify the standard \LLM softmax to accomplish three key objectives:

\begin{itemize}
    \item \textbf{Incorporate Items into the Probability Distribution:} Extend the \LLM's output distribution to encompass both the original text vocabulary $\setV$ and the item catalog $\setI$.
    \item \textbf{Reduce Computational Cost:} Avoid computing the softmax over the entire item catalog $\setI$, which is significantly larger than the text vocabulary $\setV$.
    \item \textbf{Enable Dynamic Decoding:} Allow the model to dynamically decide whether to generate a text token or an \itemid without requiring external signals or hints.
\end{itemize}

To achieve this, we define a function $c: \setI \cup \setV \rightarrow \setC$ that maps each item $i \in \setI$ and each text token $v \in \setV$ to a cluster $c(i)$ or $c(v)$ respectively, where $\setC$ is the set of all clusters.
We represent each cluster $c$ by a learnable centroid embedding $e_c \in \mathbb{R}^d$, where $\setEC$ is the matrix of all cluster centroid embeddings.

Our Two-Level softmax then operates as follows:

\textbf{First Level (Cluster Selection):} The first level models a probability distribution over all clusters $\setC$. This is achieved by computing the dot product between the decoder output $o_n$ and each cluster centroid embedding $e_c$, followed by a softmax:

\begin{equation}
\label{eq:first_pass}
    P(c|H) = \frac{\exp(\langle o_n, e_c \rangle)}{\sum_{c' \in \setC} \exp(\langle o_n, e_{c'} \rangle)}
\end{equation}

\textbf{Second Level (Item Selection):} 
The second level models the probabilities of individual tokens $w \in \setI \cup \setV$ conditioned on the token's cluster $c(w)$.
\begin{equation}
\label{eq:second_pass}
    P(w|c(w), H) =
        \frac{\exp(\langle o_n, e_w \rangle)}{\sum_{w' : c(w')=c(w)} \exp(\langle o_n,  e_{w'} \rangle)}
\end{equation}
where $e_w$ is the embedding of the target token $w$.

The overall probability of generating a token $w$ (either an item or a text token) is then given by:
\begin{equation}
\label{eq:two_pass_function}
    P(w|H) = P(c(w)|H) \cdot P(w|c(w), H)
\end{equation}

In practice, we suggest to partition the item catalog $\setI$ into approximately $\sqrt{|\setI|}$ clusters using a clustering algorithm (e.g., k-means) on pretrained item embeddings.
Each text token $v \in \setV$ can be treated as its own cluster (i.e., $c(v) = v$). In this case, for text token clusters, the centroid embedding is simply the token's embedding from $\setEV$.

\subsection{Efficient Training and Inference}

Next, we will discuss, how our proposed Two-Level softmax enables efficient training and inference.

\subsubsection{Training}
During training, the target token $w_{n+1}$ and its corresponding cluster $c(w_{n+1})$ are known (see Figure~\ref{fig:two_level_training}).
That means the cost for computing the target's probability (eq.~\ref{eq:two_pass_function}) consists of computing the cluster probability which has complexity $\O(|\setC|\cdot d)=\O((|\setV| + \sqrt{|\setI|}) \cdot d)$ and the token probability with cost $\O(\sqrt{|\setI|} \cdot d)$ in the worst case with our proposed partitioning.
The overall training cost of the proposed two level softmax is $\O((|\setV| + \sqrt{|\setI|}) \cdot d)$, which is considerably smaller than the standard softmax complexity of $O((|\setV| + |\setI|) \cdot d)$.

\subsubsection{Inference}

\label{sec:inference_optimization}

For inference, we are interested in returning the most probable tokens for a history $H$, i.e., we are interested in
\begin{align}
    \argmax_{w \in \setI \cup \setV} P(w|H)  \label{eq:inference}
\end{align}
A trivial computation of $P(w|H)$ for all tokens $w$ from two level softmax has a complexity of $O((|\setV| + |\setI|) \cdot d)$ because we would visit all clusters.

In the next paragraphs, we will discuss two approaches for fast inference.

\paragraph{Approximate Nearest Neighbor Search}

The first approach leverages existing efficient ANN search libraries like ScaNN~\cite{guoaccelerating2020} or \linebreak FAISS~\cite{johnson2019billion}. These libraries are specifically designed for fast similarity search in high-dimensional spaces and are widely used in real-time, large-scale applications, such as search engines and production recommendation systems. They can handle millions of items with extremely low latency (often sub-millisecond), making them suitable for latency-sensitive recommendation scenarios.

Generic ANN algorithms are designed to find items with maximum dot products for a set of embeddings $Z$ and a query vector $y$ by solving $\argmax_{z \in Z} \langle z, y \rangle$.
We can recast our two level softmax into such a structure to enable existing ANN algorithms:
\begin{align*}
     & \argmax_{w \in \setI \cup \setV} P(w|H) \\
    =& \argmax_{w \in \setI \cup \setV} \ln \frac{\exp(\langle o_n, e_{c(w)} \rangle)}{\sum_{c' \in \setC} \exp(\langle o_n, e_{c'} \rangle)} \frac{\exp(\langle o_n, e_w\rangle)}{\sum_{w' : c(w')=c(w)} \exp(\langle o_n, e_{w'}\rangle)}
\end{align*}
The first partition function $\sum_{c' \in \setC} \exp(\langle o_t, e_{c'}\rangle)$ is independent of any token $w$ and can be ignored when computing the maximum. Thus
\begin{align*}
     & \argmax_{w \in \setI \cup \setV} P(w|H) \\
    =& \argmax_{w \in \setI \cup \setV} \left[\langle o_n,e_{c(w)} + e_w\rangle -\ln\sum_{w' : c(w')=c(w)} \exp(\langle o_n, e_{w'}\rangle)\right] \\
    \approx& \argmax_{w \in \setI \cup \setV} \langle o_n, e_{c(w)} + e_w\rangle 
\end{align*}
Note that the last line is an approximation which is only accurate if $\ln\sum_{w' : c(w')=c} \exp(\langle o_n, e_{w'} \rangle)$ does not differ considerably between clusters.
In our experiments(Table ~\ref{tab:clustering_results}), this approximation worked well for different clustering algorithms.

To summarize, we can do inference for Two-Level softmax by applying well known ANN algorithms by indexing $e_{c(w)} + e_w, \quad \forall w \in \setV \cup \setI$ (see Figure~\ref{fig:two_level_ann}).

\paragraph{Inference Using Hierarchical Structure}

Another approach for computing the most likely token is to make use of the hierarchical structure~\cite{joulin2017bagoftricks}.
We know that the cluster probability of a token $w$, $P(c(w)|H)$ is an upper bound for the full probability of the token, i.e.:
\begin{align}
    P(w|H) = P(c(w)|H) \cdot P(w|c(w), H) \leq P(c(w)|H). \label{eq:bound}
\end{align}
This fact can be used for pruning whole clusters of items from the maximum search.
We can incrementally expand the cluster, $c^*$, with the highest probability $P(c^*|H)$ among the unvisited clusters and find the token $w^*$ with maximum probability in this cluster.
Based on the bound in equation~\ref{eq:bound}, we know that none of the other clusters $c'$ with $P(c'|H) < P(w^*|H)$ can contain a token with higher probability than $w^*$ and thus all items in these clusters can be removed from the search space.
See Figure~\ref{fig:two_level_inference} for an illustration.

\section{Properties of the Model}
\label{sec:model_properties}

Our proposed method, using the Two-Level softmax for item generation, offers several advantages in terms of efficiency and flexibility compared to existing approaches. Table~\ref{tab:model_properties} summarizes the key properties of our model compared to existing approaches. Next, we will provide more details on inference latency, item embeddings and storage.

\begin{figure*}[t]
\includegraphics[width=\linewidth]{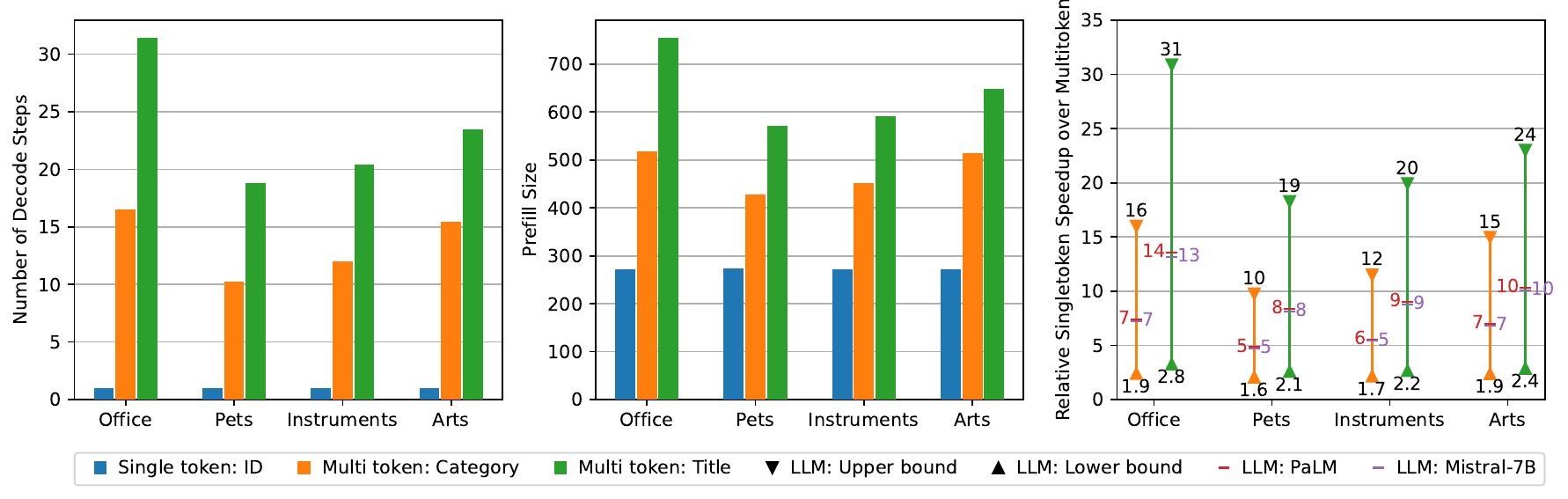}
\caption{Latency comparison: Left and middle plots show the number of decode steps / number of prefill tokens for single-token (ours) and two multi-token encoders. The actual runtime depends on the prefill and decode characteristics of the LLM inference deployment. The right plot shows the speedup of single-token (ours) over two multi-token encoders (category and title) for four different LLM deployments: PaLM from~\cite{popeefficiently2022}, Mistral-7B from~\cite{agrawal2024taming}, as well as upper and lower bounds. On the real world deployments~\cite{popeefficiently2022,agrawal2024taming}, our speedup is between 5x and 14x over multi-token approaches.}
\label{fig:latency} 
\end{figure*}

\subsection{Inference Latency}
\label{subsec:serving_latency}

Inference with an LLM consists of two parts: a prefill phase that loads the prompt and a decode phase that sequentially creates the tokens of the target item. In the decode phase, each generated token requires a full forward pass through the \LLM. Let $l_{\text{decode}}$ be the latency for decoding a single token and let $l_{\text{prefill}}(n)$ be the prefill latency where $n$ is the number of prefill tokens.
The total latency for inference depends on the token representation of the items.
If an item is represented by $m$ tokens, there are $m$ decode steps and the prompt for an item history $H$ has a length of $m\cdot |H| + \text{const}$, where \text{const} represents the other tokens of the prompt.
Thus the total latency for generating an item is $m \cdot l_{\text{decode}} + l_{\text{prefill}}(m\cdot |H| + \text{const})$. 
$m$ and $|H|$ depend on the dataset and the tokenization.
Figure \ref{fig:latency} (left) presents a comparison of the number of decoding steps required by our single-step technique ($m=1$) and multi-step baselines (which decode items as titles or categories) on the (test) datasets from our experiments in Section~\ref{sec:experiments}.
And figure \ref{fig:latency} (middle) presents a comparison of the length of prefill on the (test) datasets from our experiments.

The actual inference latency depends on the specific \LLM architecture, hardware and backbone LLM implementation, in particular, on the deployment characteristics $l_{\text{decode}}$ and $l_{\text{prefill}}(n)$.
For example, with the characteristics of Mistral-7B~\cite{agrawal2024taming}, the overall prefill+decode latency for our method is 35ms + 20ms (55ms) and for our baselines using "title" \multitoken encoding is 75ms + 400ms (475ms), improving inference efficiency by 8x. For PALM~\cite{popeefficiently2022}, our method is 48ms + 20ms (68ms) and our baselines using "title" \multitoken encoding is 96ms + 580ms (676ms).
Figure~\ref{fig:latency} (right) shows the speedup of our single-step technique over multi-step techniques for various datasets with the Mistral-7B~\cite{agrawal2024taming} and PaLM~\cite{popeefficiently2022} characteristics.
Finally, we provide an analysis for \emph{any} possible LLM deployment for which $l_{\text{prefill}}(n)=n\cdot l_{\text{prefill}}$, i.e., any LLM that has linear prefill costs.
In this case, the speedup is $\frac{m\cdot l_{\text{decode}} + (m\cdot |H| + \text{const})\cdot l_{\text{prefill}}}{l_{\text{decode}} + (|H| + \text{const})\cdot l_{\text{prefill}}} $.
For any latency characteristics ($l_{\text{decode}}$ and $l_{\text{prefill}}$), this function is upper-bounded by $m$ and lower-bounded by $\frac{m\cdot |H| + \text{const}}{|H| + \text{const}}$.
Figure~\ref{fig:latency} (right) shows these upper and lower bounds for the speedup on our datasets.

\subsection{Flexible Item Embedding Dimension}

Our approach uses a separate embedding table $\setEI$ for item representations, which is distinct from the \LLM's text token embedding table $\setEV$. This separation offers flexibility in choosing the dimensionality of the item embeddings. While the \LLM's text embeddings typically have a fixed dimension (e.g., the $d$ in the PaLM 2 model), we can choose a different dimension $k$ for the item embeddings. In our experiments, we use an item embedding dimension of $k=512$. This flexibility allows us to optimize the item embeddings for the recommendation task without being constrained by the pre-trained \LLM's architecture.

\subsection{Storage Considerations}

Our method requires storing the item embedding table $\setEI$ in addition to the \LLM parameters. This does introduce additional storage requirements compared to methods that solely rely on the \LLM's internal vocabulary. 
For example, a model with 1 million items and an item embedding dimension of $k=500$ would need to store an additional $0.5$ billion item parameters.
However, the storage overhead is manageable in practice and comparable to that of traditional recommendation models that also store item embeddings. If needed techniques \cite{hashing_1, hashing_2} exist to optimize this.

\section{Experiments}
\label{sec:experiments}

Our experiments investigate the prediction quality of our proposed \singlestep decoding methods and compare them to other approaches: \LLM based recommenders with \multistep decoding and non-\LLM recommenders.
We also investigate the impact of individual components of our \singlestep algorithms, specifically the inference algorithm, clustering method and the underlying \LLM.

\begin{table}[t]
\centering
\begin{tabular}{|l|cccc|}
\hline
\textbf{Dataset} & \textbf{\# of Users} & \textbf{\# of Items} & \textbf{Items/} & \textbf{Purchases/} \\
 & & & \textbf{User} & \textbf{Item} \\
\hline
Office & 44,736 & 27,482 & 7.87 & 12.81 \\
Pets & 43,135 & 37,712 & 8.82 & 10.09 \\
Instruments & 25,577 & 10,599 & 8.39 & 20.24 \\
Arts & 47,197 & 22,828 & 8.72 & 18.02 \\
\hline
\end{tabular}
\caption{Statistics of the Amazon Review Datasets}
\label{tab:dataset_stats}
\end{table}

\subsection{Setup}
\subsubsection{Dataset and Preprocessing}

We use the widely-used Amazon Review dataset (2018 version)~\cite{ni2019justifying}, specifically 4-core subsets in Table ~\ref{tab:dataset_stats}, following the preprocessing steps outlined in CALRec~\cite{CALRec}. User interactions are sorted chronologically by timestamp. We adopt the standard leave-one-out evaluation protocol: for each user, the last interaction is used for testing, the second-to-last for validation, and the remaining interactions for training. Like the baselines, our method also utilizes item features: title, category, brand, and price.

The Amazon Review dataset is popular and widely\cite{CALRec, recformer, tiger2023} used in the field of LLM-based recommendations due to the rich textual metadata about items present in this dataset.

\subsubsection{Model Configuration}
We use a batch size of 512, packing multiple examples into each batch and using masking to prevent attention across example boundaries. We use a learning rate of 5e-3, a cosine learning rate decay, and a weight decay of 1e-5. We train for a maximum of 50,000 steps. When using the larger backbone \LLM model, PaLM 2 XS, we update weight decay to 1e-2. Labels are replaced with a placeholder in the input to the model and causal mask is used. We use a randomly initialized item embedding of dimension $k=512$. This embedding is trained along with the LLM for the final task. For our Two-Level softmax, the clusters are pre-computed as a pre-processing step. Recall that \ref{sec:training_paradigm} enabled us to use only \itemid representation for test set.

\begin{table*}[htbp]
\centering
\begin{tabular}{|l|l|ccccc|cc|ccc|}
\hline
\multicolumn{2}{|c|}{} & \multicolumn{5}{c|}{\textbf{Non-LLM Methods}} & \multicolumn{5}{c|}{\textbf{LLM Based Methods}}  \\

\multicolumn{2}{|c|}{} & \multicolumn{5}{c|}{} & \multicolumn{2}{c|}{Multi-Step Decoding} & \multicolumn{3}{c|}{Single-Step Decoding (Ours)} \\
\hline
Dataset & Metric & BERT4Rec & SASRec & FDSA & S3-Rec & UniSRec & RecFormer & CALRec & Full Softmax & Structure & ANN \\
\hline
Office & Recall@1 & 0.0403 & 0.0134 & 0.0547 & 0.0291 & 0.0250 & 0.0328 & 0.0761 & \textbf{0.0839} & \underline{0.0814} & 0.0761 \\
       & Recall@10 & 0.0709 & 0.1033 & 0.0982 & 0.0941 & 0.1220 & 0.1063 & 0.1213 & \textbf{0.1330} & \underline{0.1234} & 0.1213 \\
       & NDCG@10 & 0.0545 & 0.0602 & 0.0749 & 0.0607 & 0.0713 & 0.0687 & 0.0976 & \textbf{0.1057} & \underline{0.1004} & 0.0970 \\
       & MRR & 0.0520 & 0.0499 & 0.0710 & 0.0535 & 0.0597 & 0.0603 & 0.0925 & \textbf{0.0973} & \underline{0.0932} & 0.0887 \\
\hline
Pets & Recall@1 & 0.0276 & 0.0093 & 0.0394 & 0.0162 & 0.0202 & 0.0377 & 0.0570 & \textbf{0.0643} & \underline{0.0634} & 0.0625 \\
     & Recall@10 & 0.0499 & 0.0773 & 0.0710 & 0.0647 & 0.0970 & 0.0826 & 0.0937 & \textbf{0.1104} & \underline{0.1094} & 0.1032 \\
     & NDCG@10 & 0.0376 & 0.0448 & 0.0537 & 0.0392 & 0.0574 & 0.0590 & 0.0736 & \textbf{0.0839} & \underline{0.0819} & 0.0798 \\
     & MRR & 0.0365 & 0.0382 & 0.0522 & 0.0349 & 0.0503 & 0.0549 & 0.0696 & \textbf{0.0760} & \underline{0.0741} & 0.0730 \\
\hline
Instruments & Recall@1 & 0.0494 & 0.0158 & 0.0574 & 0.0202 & 0.0237 & 0.0266 & \textbf{0.0718} & \textbf{0.0718} & \underline{0.0688} & \underline{0.0688} \\
      & Recall@10 & 0.0915 & 0.1130 & 0.1092 & 0.1048 & 0.1255 & 0.0940 & 0.1158 & \textbf{0.1297} & \underline{0.1289} & 0.1225 \\
      & NDCG@10 & 0.0680 & 0.0623 & 0.0796 & 0.0606 & 0.0709 & 0.0596 & 0.0909 & \textbf{0.0958} & \underline{0.0919} & 0.0913 \\
      & MRR & 0.0658 & 0.0528 & 0.0765 & 0.0526 & 0.0613 & 0.0535 & \underline{0.0864} & \textbf{0.0868} & 0.0814 & 0.0818 \\
\hline
Arts & Recall@1 & 0.0347 & 0.0147 & 0.0460 & 0.0216 & 0.0213 & 0.0279 & \textbf{0.0636} & \underline{0.0626} & 0.0600 & 0.0590 \\
     & Recall@10 & 0.0799 & 0.1118 & 0.1061 & 0.1052 & \textbf{0.1320} & 0.1095 & 0.1140 & \underline{0.1229} & 0.1147 & 0.1125 \\
     & NDCG@10 & 0.0547 & 0.0619 & 0.0726 & 0.0602 & 0.0729 & 0.0662 & \underline{0.0864} & \textbf{0.0884} & \underline{0.0864} & 0.0851 \\
     & MRR & 0.0513 & 0.0519 & 0.0680 & 0.0520 & 0.0615 & 0.0582 & \textbf{0.0815} & \underline{0.0780} & 0.0758 & 0.0753 \\
\hline
\end{tabular}
\caption{Comparison of our efficient LLM based recommender to existing non-LLM recommenders and \LLM methods. The best results are highlighted in bold, the second best are underlined. Results for baselines are taken from CALRec~\cite{CALRec}}
\label{tab:main_results}
\end{table*}

\begin{table*}[htbp]
\centering
\begin{tabular}{|l|l|ccc|ccc|}
\hline
\multicolumn{2}{|c|}{} & \multicolumn{3}{c|}{\textbf{Structure}} & \multicolumn{3}{c|}{\textbf{ANN}} \\
\hline
Dataset & Metric & Kmeans & Frequency & Random & Kmeans & Frequency & Random \\
\hline
Office & Recall@1 & 0.0781 & \textbf{0.0814} & \underline{0.0791} & \textbf{0.0761} & \underline{0.0738} & 0.0719 \\
       & Recall@10 & \textbf{0.1244} & \underline{0.1234} & 0.1207 & \underline{0.1213} & \textbf{0.1250} & 0.1201 \\
       & NDCG@10 & \underline{0.0986} & \textbf{0.1004} & 0.0978 & \textbf{0.0970} & \underline{0.0964} & 0.0939 \\
       & MRR & 0.0907 & \textbf{0.0932} & \underline{0.0908} & \textbf{0.0887} & \underline{0.0884} & 0.0858 \\
\hline
Pets & Recall@1 & \textbf{0.0634} & 0.0610 & \underline{0.0615} & \textbf{0.0625} & 0.0566 & \underline{0.0598} \\
     & Recall@10 & \underline{0.1094} & 0.1043 & \textbf{0.1095} & \underline{0.1032} & 0.1017 & \textbf{0.1074} \\
     & NDCG@10 & \textbf{0.0819} & 0.0804 & \underline{0.0814} & \underline{0.0798} & 0.0764 & \textbf{0.0800} \\
     & MRR & \textbf{0.0741} & 0.0731 & \underline{0.0735} & \textbf{0.0730} & 0.0687 & \underline{0.0717} \\
\hline
Instruments & Recall@1 & \underline{0.0680} & 0.0679 & \textbf{0.0688} & 0.0665 & \underline{0.0667} &  \textbf{0.0688} \\
      & Recall@10 & \underline{0.1207} & 0.1204 & \textbf{0.1289} & 0.1170 &  \textbf{0.1231} & \underline{0.1225} \\
      & NDCG@10 & \underline{0.0905} & 0.0903 & \textbf{0.0919} & 0.0880 & \underline{0.0884} &  \textbf{0.0913} \\
      & MRR & \underline{0.0813} & 0.0811 & \textbf{0.0814} & 0.0792 & \underline{0.0808} &  \textbf{0.0818} \\
\hline
Arts & Recall@1 & \textbf{0.0600} & \underline{0.0568} & 0.0542 & \textbf{0.0590} & \underline{0.0565} & 0.0543 \\
     & Recall@10 & \textbf{0.1147} & \underline{0.1024} & 0.1002 & \textbf{0.1125} & 0.1011 & \underline{0.1097} \\
     & NDCG@10 & \textbf{0.0864} & \underline{0.0842} & 0.0805 & \textbf{0.0851} & \underline{0.0841} & 0.0800 \\
     & MRR & \textbf{0.0758} & \underline{0.0745} & 0.0715 & \textbf{0.0753} & \underline{0.0736} & 0.0710 \\
\hline
\end{tabular}
\caption{Ablations on clustering methods (Kmeans, Frequency and Random) for Two-Level softmax when using ANN and Structure inference.}
\label{tab:clustering_results}
\end{table*}

\begin{table}
\centering
\begin{tabular}{|l|l|cc|}
\hline
\textbf{Dataset} & \textbf{Metric} & \textbf{PaLM 2 XXS} & \textbf{PaLM 2 XS} \\
\hline
Office & Recall@1 & 0.0761 & \textbf{0.0780} \\
       & Recall@10 & 0.1213 & \textbf{0.1290} \\
       & NDCG@10 & 0.0970 & \textbf{0.1011} \\
       & MRR & 0.0887 & \textbf{0.0903} \\
\hline
\end{tabular}
\caption{Ablations on \LLM size: PALM 2 XXS 1B parameters and XS 8B parameters with Two-Level softmax ANN inference.}
\label{tab:modelsize_results}
\end{table}

\subsubsection{Evaluation Metrics}

Following standard practice in sequential recommendation, we evaluate performance using NDCG@K (Normalized Discounted Cumulative Gain), Recall@K and MRR (Mean Reciprocal Rank). We compute these metrics by scoring the full item catalog for each user in the test set and averaging the results.

\subsubsection{Baselines}

We compare our proposed \singlestep decoding method against the following baselines:

\begin{itemize}
\item \textbf{SASRec}~\cite{kang2018self}: Unidirectional Transformer for sequential recommendation, using item IDs.
\item \textbf{BERT4Rec}~\cite{sun2019bert4rec}: Bidirectional Transformer with masked language modeling, using item IDs.
\item \textbf{FDSA}~\cite{FDSA}: Two self-attention blocks to address ID sequence and attribute feature sequence, respectively, and fuses their final representations for next-item prediction.
\item \textbf{S$^3$-Rec}~\cite{s3rec}: Contrastive learning on item IDs and attributes, with pretraining and fine-tuning of encoder-only transformer network.
\item \textbf{UniSRec}~\cite{unisrec}: Encodes item text with BERT (plus adapter) and uses a sequence model, where an item embedding is the sum of its ID and text representations.
\item \textbf{Recformer}~\cite{recformer}: Longformer with token type and position embeddings, pretrained with MLM and contrastive loss. They fine-tune a pretrained checkpoint and represent items as text sentences.
\item \textbf{CALRec}~\cite{CALRec}: Uses \multistep decoding with item metadata, specialized beam search, and contrastive losses. CALRec and our method, use PaLM 2-XXS~\cite{anil2023palm} as the backbone pre-trained \LLM for a fair comparison, focusing on the impact of the decoding strategy.
\end{itemize}

\subsection{Results}
\label{sec:results}

Table~\ref{tab:main_results} presents the main results of our experiments, comparing our methods Full Softmax (Section~\ref{subsec:fullsoftmax}), Two-Level ANN (Section~\ref{sec:inference_optimization}), and Two-Level Structure (Section~\ref{sec:inference_optimization}) against the baselines on the Amazon dataset.

\subsubsection{\Singlestep decoding methods}
In Table~\ref{tab:main_results}, we compare our \singlestep \singletoken techniques with LLM-based \multistep \multitoken methods CALRec and Recformer and demonstrate that Full Softmax outperforms them with highest quality. This outperformance suggests that the more direct optimization of item decoding, leads to better performance in this setting. Our proposed method (Two-Level Softmax with clustering) can be used with our two proposed efficient inference algorithms (Section~\ref{sec:inference_optimization}): Structure and ANN. Structure inference utilizes the hierarchical structure, while ANN inference uses Approximate Nearest Neighbor search to approximate this hierarchical structure. Both inference techniques achieve results comparable to the strong \LLM-based baselines, demonstrating that these approaches do not sacrifice recommendation quality for efficiency. Overall, all three approaches significantly outperform the traditional non-\LLM recommendation methods (SASRec, BERT4Rec, FDSA, S3-Rec, UniSRec) on the sequential recommendation task. For the datasets used, we observe approximately a 10\% improvement in training speed when using Two-Level softmax as opposed to Full softmax. In scenarios where training with Full Softmax is manageable, we recommend Full Softmax for optimal performance. However, when Full Softmax becomes impractical due to computational constraints, we recommend ANN or Structure inference based on the desired level of efficiency.

\subsubsection{Ablation of inference algorithm}
We ablate the performance of our two proposed inference methods (Section~\ref{sec:inference_optimization}): Structure and ANN in Table~\ref{tab:main_results}. ANN represents our simplest setup, approximating the term, $\ln\sum_{w' : c(w')=c} \exp(\langle o_n, e_{w'} \rangle)$. While Structure inference generally performs better than ANN inference, the quality of ANN remains competitive to existing \multistep baselines.

\subsubsection{Ablation of clustering method}
In Table~\ref{tab:clustering_results}, we ablate the choice of clustering method used in our Two-Level softmax. We compare three different clustering techniques: Kmeans, Frequency, and Random. All three techniques are performed as a pre-processing step before the LLM is trained. For Kmeans, we apply K-means clustering on item embeddings generated using an open-source matrix factorization model from \cite{rendlerevisiting2022} and use the resulting clusters to partition the items into approximately $\sqrt{|\setI|}$ clusters. For Frequency, we count the frequency of item occurrences in the training set and group items with similar occurrence counts into the same cluster. For Random, we randomly assign items to clusters. For ANN inference, the choice of clustering method appears to have a less significant impact, with all three methods performing relatively similarly. Interestingly, Random clustering performs well in this setting. We hypothesize that this might be because random clustering makes the approximation term in ANN, $\ln\sum_{w' : c(w')=c} \exp(\langle o_n, e_{w'} \rangle)$, more uniform across different clusters. Performance of Kmeans clustering is notably better with Structure inference than with ANN based inference across all metrics. For the Office dataset, Structure inference with Frequency-based clustering achieves the best performance.

\subsubsection{Ablation of underlying \LLM}
Although we present most of our results and comparisons with baselines using PaLM2 XXS for a fair comparison to the baseline method CALRec, we also investigate increasing the LLM size from PaLM 2-XXS(1B) to PaLM 2-XS(8B) in Table~\ref{tab:modelsize_results} and this further improves the recommendation quality, suggesting that our approach can benefit from more powerful language models.

\subsubsection{Ablation of training paradigm}
CALRec uses multiple training stages to first align the LLM with item domain. However we find that with our training paradigm in Section \ref{sec:training_paradigm} we can eliminate these stages and achieve quality. We also experimented with item pretraining stages using \cite{rendlerevisiting2022} to learn collaborative \itemid embeddings and \cite{gecko} to learn semantic embeddings. With Section \ref{sec:training_paradigm} we were able to achieve neutral quality with and without these pretraining stages and hence we do not include these results in the paper.

\subsubsection{Summary} The key findings of our study are:
\begin{itemize}
\item Improved quality: Our efficient single-stage training approach with Two-Level Softmax, achieves recommendation quality comparable to, and in some cases exceeding, multi-step multi-token LLM-based baselines, and significantly outperforming traditional non-LLM methods.
\item Clustering Method Impact: While the clustering method can have some impact, especially with Structure inference, even simple methods like Random clustering can achieve competitive performance, particularly with ANN inference.
\item Impact of LLM Size: Increasing the LLM size improves recommendation quality, highlighting the importance of our efficient method.
\item Inference latency: The reduction in input prompt length improves prefill latency, and the reduction in decode steps enhances decode latency (Figure \ref{fig:latency} for Mistral and PALM).
\end{itemize}

\section{Conclusion}
\label{sec:conclusion}

This paper addresses the critical latency bottleneck in \LLM-based recommender systems which is a crucial limitation for applying LLMs in practical real-time recommendation. Contrary to prevailing methods, we demonstrate the feasibility and superiority of treating items (or any other ids) as first-class citizens of the LLM rather than forcing them through the default LLM text tokenization framework as \multitoken. We identify and optimize the critical softmax bottleneck over large item vocab arising from this approach and introduce a novel \singlestep item decoding enabled by two-level softmax. Combined with our unique, efficient end-to-end training paradigm that eliminates the need for separate pre-training stages, our approach achieves higher quality compared to existing multi-stage and \multistep approaches on the Amazon product recommendation benchmarks while significantly reducing inference latency. Our key contributions are (1)~a novel perspective on prefill-decode latency of existing approaches, (2)~presenting a highly effective \singletoken approach, and (3)~offering a simplified, single-stage training algorithm that achieves state-of-the-art quality. This work demonstrates the potential for efficiently bridging the gap between powerful \LLM capabilities and the practical constraints of large-scale, real-time recommendation, potentially inspiring a new direction for future research in LLM-based recommendation. Future research directions include exploring the integration of external embeddings, use of semantic understanding for clustering and use of multi-modal item representations in our training paradigm.

\section{GenAI Usage Disclosure}
GenAI was not used to create or edit this manuscript. All sections and code have been created and verified by authors.


\bibliographystyle{ACM-Reference-Format}
\bibliography{references}

\appendix

\end{document}